\documentclass{bioinfo}
\copyrightyear{2011}
\pubyear{2011}

\usepackage{amsmath,  amssymb}

\newtheorem{theorem}{Theorem}[section]

\newcommand\lf{\mbox{Leaf}}

\begin{document}
\title[Non-binary tree Reconciliation]{Reconciliation of Gene and Species Trees With Polytomies}

\author[Zheng \textit{et~al}]{Yu Zheng\,$^{1}$, Taoyang Wu\,$^{1}$, Louxin Zhang\,$^{1}$
\footnote{to whom correspondence should be addressed}}
 \address{$^{1}$Department of Mathematics,
        Nat'l University of Singapore,
        10 Lower Kent Ridge, S'pore 119076
        }

\history{Received on XXXXX; revised on XXXXX; accepted on XXXXX}

\editor{Associate Editor: XXXXXXX}

\maketitle

\begin{abstract}

\section{Motivation:} Millions of genes in the modern species belong to
 only thousands of gene families. 
Genes duplicate and  are lost during evolution. 
A gene family
includes instances of the same gene in different species
and duplicate genes in the same species. 
Two genes in different species are ortholog if their common ancestor lies in
the most recent common ancestor of the species.
Because of complex gene evolutionary history, ortholog identification is a basic but
 difficult
task in comparative
 genomics. A key method for the task is to use an explicit model of
the evolutionary history of the genes being studied, called the gene
(family) tree. It compares the gene tree with the evolutionary
history of the species in which the genes reside, called the species
tree, using the procedure known as tree reconciliation. Reconciling
binary gene and specific trees is simple. However, 
tree reconciliation presents challenging problems when species trees are not binary in practice. 
Here, arbitrary  gene and species tree reconciliation is studied in a binary
refinement model.
\section{Results:} The problem of reconciling gene and
species trees is proved NP-hard when species tree is not binary  even for the duplication cost.
 We
then present the first efficient method for reconciling a non-binary
gene tree and a non-binary species tree. It attempts to find a binary
refinement of the given gene and species trees that minimizes the given 
reconciliation cost if they are not binary.
 Our algorithms have been implemented into a software to
  support quick automated analysis of large data sets.
\section{Availability:} The  program, together with the source code, is  available at
its online server http://phylotoo.appspot.com.
\section{Contact:}
\href{yu_zheng@nus.edu.sg}{yu\_zheng@nus.edu.sg} or matzlx@nus.edu.sg

\end{abstract}

\section{Introduction}

Millions of genes in the modern species are not completely
independent of one another;  they belong to only 
thousands  of gene families instead. A gene family includes instances of
the same gene in different species  and duplicate genes
in the same species.  
Orthology refers to a specific relationship between
homologous characters that arose by speciation at their most recent
point of origin (Fitch, 1970).  Two genes in different species are
ortholog if they arose by speciation in the most recent common ancestor of the
species. 
Orthologous genes  tend to retain similar
biological functions, whereas non-orthologs  often diverge over
time to perform different functions via subfunctionalization and
neofunctionalization.
 Ortholog identification is the first task of
 almost every comparative genomic study since
 orthologs
  are used to infer the pattern of gene gain and loss,  the mode of signaling pathway evolution, and 
the correspondence between genotype and phenotype.

Genes are gained through duplication and horizontal gene transfer and lost
via deletion and pesudogenization throughout evolution.
 Identifying
orthologs is essentially to find out how genes evolved. 
Since  past evolutionary 
events
cannot be observed directly,  we have to infer these events from the gene
sequences available today. Therefore,  ortholog
identification is never an easy task.

A key method for ortholog identification is to use an explicit model
of the evolutionary history of the genes subject to study, in the
form of a gene family tree. It compares the
 gene tree with the evolutionary history of the
species the genes reside in -- the species tree -- using the
procedure known as tree reconciliation (Goodman \textit{et al.},
1979; Page, 1994). The rationale underlying this approach is that,
by parsimony principle, the smallest number of evolutionary events
is likely to reflect the evolution of a gene family. Gene tree and
species tree reconciliation formalizes the following intuition: If
the offspring of a node in a gene tree is distributed in the same
set of species as that of a direct descendant, then the node
corresponds to a duplication. Different reconciliation algorithms
for inferring gene duplication, gene loss,  and other events have
been developed (Arvestad \textit{et al.}, 2004;
 Berglund \textit{et al.}, 2006;
Chang and Eulenstein, 2006; Durand \textit{et al.}, 2005; Ma
\textit{et al.}, 2000;  Vernot \textit{et al.}, 2008). The tree
reconciliation approach is less prone to error than heuristic
sequence-match methods particularly in the situation when gene loss
events are not rare (Kristensen \textit{et al.}, 2011).

The concept of tree reconciliation is rather simple.   Standard
reconciliation map from a binary gene tree to a binary species tree
is linear-time computable (Chen \textit{et al.}, 2000;
 Zhang, 1997; Zmasek and Eddy, 2001).
 However, tree reconciliation presents challenging problems when the input species tree
is not binary in practice. A gene (family) tree is reconstructed from the sequences
of its family members. When a maximum likelihood or Bayesian method is
used for the purpose, the output gene tree often contains non-binary
nodes. Such nodes are called soft polytomies (Maddison, 1989)
because the true pattern of gene divergence is binary (Hudson,
1990), but there is not enough signal in the data to time the true
diverging events. On top of ambiguity in gene tree, there are also
uncertainties in a species tree. The NCBI taxonomy database and
other reference species trees are often non-binary due to unsolved
species diverging order, for example in the case of eukaryote
evolution (Koonin, 2010). Reconciling non-binary gene and species
trees is a daunting task. The standard reconciliation used for
binary gene and species trees will not produce correct gene
evolution history when applied to non-binary species trees. The
complexity of the general reconciliation problem is unknown
(Eulenstein \textit{et al.}, 2010).  Notung, one of the best packages
for tree reconciliation, requests that  one of the two reconciled
trees has to be binary (Durand \textit{et al.}, 2005; Vernot
\textit{et al.}, 2008).

{\bf Related work and our contribution}~~In this work, we focus on
the two issues mentioned above.   Recently, tree reconciliation has
been studied in different models and for different types of gene
trees. For a binary species tree and a non-binary gene tree, the
reconciliation problem can be solved  via a dynamic programming
approach in polynomial time (Chang and Eulenstein, 2006; Durand
\textit{et al.}, 2005). The duplication/loss cost is used in (Chang
and Eulenstein, 2006), whereas the weighted sum of gene duplication
and loss costs is used in (Durand
\textit{et al.}, 2005).

Resolving non-binary gene tree nodes was also independently studied
 for arbitrary
species trees in (Berglund \textit{et al.}, 2006), where the
optimality criteria used is minimization of duplications and
subsequently loss events.
 A heuristic search algorithm was proposed to compute the
number of duplications necessary for resolving a non-binary gene
tree node. The gene loss cost is computed subsequently  after
duplications are inferred. Because of its heuristic nature, the
method might stop before a solution with the best reconciliation
score is found and hence sometimes overestimates 
the number  of loss events.

Conversely, reconciliation with non-binary species trees is much
harder and less studied. Vernot et al. (2008) proposed two types of
duplications for studying this problem: {\it required}  and {\it
conditional} duplications. The latter is used to indicates that a
disagreement between a gene tree node and a non-binary species tree
node is detected, but it is impossible to determine whether gene
duplication or other events such as incomplete lineage sorting are
responsible for the disagreement. These two types of duplications
are efficiently computable.

In this work, we study the general reconciliation problem by finding
 binary refinements of the given gene tree and species tree with the
minimum reconciliation cost over all possible pairs of such binary
refinements (see Section 2 for the definition of binary refinement).
Such a reconciliation model is first formulated in (Eulenstein
\textit{et al.}, 2010). We prove that the reconciliation problem
 is NP-hard even for a binary gene
tree and a non-binary species tree, solving an open question raised
in the reconciliation study (Eulenstein \textit{et al.}, 2010).
 We then propose a two-stage method for
reconciling arbitrary gene and species trees. The first stage of the
method is based on a novel algorithm for resolving non-binary
species tree nodes using structural information of the input gene
tree. The algorithm is simple, but very efficient as shown by our
validation test. The second stage of our method uses a new linear
time algorithm for resolving a non-binary gene tree with a binary
species tree. It is a natural extension of the standard
reconciliation procedure from binary gene trees to non-binary gene
trees.

To our knowledge, no formal algorithm for reconciling two non-binary
trees has been reported.
 Our approach has been implemented in a software package,
whose online server is
 on http://phylotoo.appspot.com.

\section{Algorithms and Methods}

\subsection{Basic concepts and notations}

{\bf Gene trees and species trees}~~In this study, we focus on
rooted gene trees and species trees. A rooted tree $T$ is a graph in
which there is exactly a distinguished node, called the root, and
there is a unique path from the root to any other node. We define
 a partial order 
$\leq_T$ on the
node set of $T$: $v\leq_T u$ if and only if $u$ is in the
path from  the root to $v$. Furthermore, we define $v<_T
u$ if and only if $v\leq_T u$ and $v \neq u$.
 We shall write $\leq$ and $<$
 whenever no confusion will arise after the subscript $T$ is dropped.

Obviously, the root is the maximum element under $\leq$ in $T$. The
minimal elements under $\leq$ are called the {\it leaves} of $T$.
The leaf set is denoted by $\lf(T)$. Non-leaf nodes are  called {\it
internal nodes}.  The set of the internal nodes of $T$ is denoted by
$V(T)$. For each $u\in V(T)$, all the nodes $v$ satisfying that $v\leq u$
form a subtree rooted at $u$, denoted by $T(u)$. For any $v\in T(u)$,  $v$  is called
a {\it descendant} of $u$ or $u$ an {\it ancestor} of $v$  if $v\neq u$;
$v$ is called  a
{\it child} of $u$ if there is no $u'$ such that $v< u'<
u$. 
A tree node is {\it binary} if it has exactly two children; it is
{\it non-binary} otherwise.  $T$ is  {\it binary} if all the internal
nodes are binary in $T$  and {\it non-binary} otherwise.

 For a nonempty $I\subseteq V(T)\cup \lf(T)$, $x$ is a
common ancestor of $I$ if it is an ancestor of
 every  node $y\in I$; a common ancestor is the {\it least common ancestor} (lca) of $I$
if none of its children
 is  a common ancestor of $I$. The lca of $I$ is written $\mbox{lca}(I)$.

A gene  or species tree is a rooted tree with labeled leaves.
For a gene or species tree $T$, we shall use $L(T)$ to denote the set of leaf labels found in $T$.
Each species tree leaf has a modern species as its label.
A gene tree is built from the DNA or protein sequences of  a gene
family. In a gene tree $G$, each leaf represents a member of the
gene family. In the study of gene tree and species tree reconciliation,  
a gene tree leaf is labeled with the species in which it resides.
  Since the gene family often includes  duplicate genes in
  the same species,  a gene tree is often not uniquely leaf labeled. 
For each
 $g\in V(G)$, we use $L(g)$ to denote the set of the leaf labels in
the subtree $G(g)$. Because of duplicate genes in a gene family,  $L(g)$ and $L(g')$ can be equal for
different $g$ and $g'$ in $G$. \vspace{1em}

{\bf Tree reconciliation}~~ Consider a species tree $S$  and a gene
tree $G$ of a gene family whose members are found in the species in
$L(S)$.  A {\it reconciliation} $f$ between $G$ and $S$ is a map
from the gene tree nodes to the species tree nodes having the
following
properties:\\

\begin{quote}
\noindent
  (i) (Leaf-preserving) For any $x\in \lf(G)$, $f(x)\in \lf(S)$ and
 has the same label as $x$.

 \noindent (ii) (Order-preserving) For any gene tree nodes $g$ and $g'$
 such that  $g'\leq_G g$, $f(g')\leq _S f(g)$.
\end{quote}

Furthermore,  the {\it lca reconciliation} $\lambda$  maps $u$ to\\
$\mbox{lca}\left(\{\lambda(x): x\in \lf(G(u))\}\right)$. It is easy
to see that for any $g\in V(G)$ with $k$ children $g_1, g_2, \cdots,
g_k$,
 $\lambda(g)=\mbox{lca}(\{\lambda(g_i): i\leq k\})$.
 Note that $\lambda$ is a special reconciliation between $G$ and
 $S$.
The lca reconciliation is the minimum one in the sense that, for any
reconciliation $f$,  $\lambda(u)\leq _S f(u)$ for every $u\in V(G)$.
\vspace{1em}

{\bf Tree refinement}~~In graph theory, an edge contraction is an
operation which removes an edge from a graph while simultaneously
merging together the two vertices previously connected through the
edge.  For two gene trees $G$ and $G'$, $G$ is said to refine $G'$
if $G'$ can be obtained from contracting edges in $G$. If $G$
refines $G'$, we can map each node of $G'$ to a unique node in $G$
such that the ancestral relationship is preserved. The species tree
refinement can be defined similarly. \vspace{1em}

{\bf General Reconciliation Problem}~~ In this paper, we shall study
tree reconciliation through the binary refinement of non-binary gene  and species
trees (Eulenstein \textit{et al.}, 2010):  Given a gene tree $G$,  a
species tree $S$, and a reconciliation cost function,  find a binary
refinement $G'$ of $G$ and a binary refinement $S'$ of $S$ such that
the reconciliation of $G'$ and $S'$ has the minimum reconciliation
cost over all such refinements. We shall work with the gene
duplication cost, the gene loss cost, or the weighted sum of these
two costs. Due to space limitation, these cost models for binary
gene tree and binary species tree reconciliation will not be defined
here. The readers are referred to (Eulenstein \textit{et al.}, 2010;
Ma \textit{et al.}, 2000) for the definitions.

 \subsection{NP-Hardness of the General Reconciliation Problem}

Unfortunately, the general reconciliation problem is computationally
hard for non-binary species trees. More specifically, we prove it
NP-hard via a reduction from the problem of constructing a species
tree from a set of gene trees. The complexity of the latter has been 
investigated in (Ma \textit{et al.}, 2000; Bansal and Shamir, 2010).
The full proof can be found in the Section A of the supplementary
document.

\begin{theorem}
Gene tree and species tree reconciliation via binary refinement is NP-hard
 for non-binary species trees even for the duplication cost.
\end{theorem}

\begin{figure}[!th]
\begin{center}
\includegraphics[width=0.95\columnwidth]{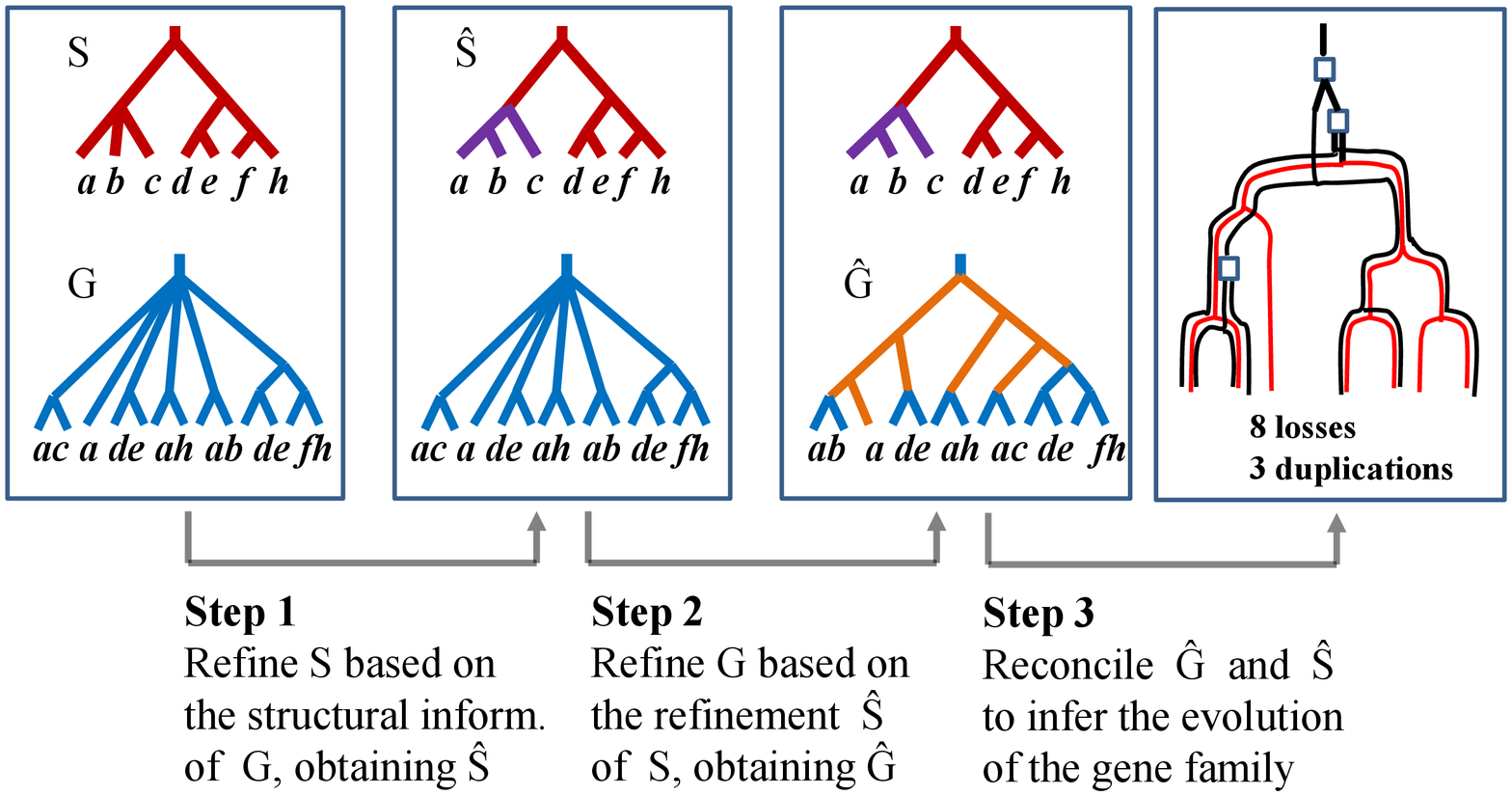}
\end{center}
\caption{A schematic view of our method for reconciling a non-binary
gene tree $G$ of a gene family  and a non-binary species tree $S$.}
\label{Fig1_2StageM}
\end{figure}

\subsection{A Heuristic Reconciliation Method}

Since the general reconciliation problem is NP-hard, it unlikely has
a polynomial-time algorithm. An efficient heuristic method for it is
developed here.

As illustrated in Fig.~\ref{Fig1_2StageM}, the method consists of
three steps. Given an arbitrary gene tree $G$ of a gene family and the
containing species tree $S$, our method first computes a binary
refinement $\hat{S}$ of $S$ using the structural information of $G$;
it then computes a binary refinement $\hat{G}$ of $G$ based on
$\hat{S}$ in the second step; finally, it outputs a hypothetical duplication
history of the gene family by reconciling $\hat{G}$ and $\hat{S}$.

Reconciliation of a binary gene tree and a binary species tree is
well studied. We shall only describe the detail of the first and
second steps in the rest of this section.

\subsection{Step One: Resolve Non-binary Species Tree Nodes}

Our algorithm for resolving non-binary species tree nodes is motivated by
the following facts. Recall that the lca reconciliation map is
denoted by $\lambda$.
Assume the input gene and species trees be $G$ and $S$, respectively, where 
$G$  may not be binary. We resolves the non-binary nodes in $S$ one by one. 

 Consider a non-binary node $s\in S$ having children
 $s_1, s_2, \cdots, s_{n(s)}$, where $n(s)\geq 3$.
We define the preimage set
 $$\mbox{Pre}(s)=\{g \in V(G): \lambda(g)=s\}$$
of $s$ under $\lambda$.
Then, $\mbox{Pre}(s)$ has the following properties:

\begin{itemize}
\item For each $g\in \mbox{Pre}(s)$, there are at least two children $s_i$ and $s_j$
 of $s$ such that
   $$L(g) \cap L(s_i) \neq \phi,\; L(g)\cap L(s_j) \neq \phi.$$
In other words, some descendants of $g$ are found in  modern
species evolving from $s_i$, whereas some other descendants of $g$ are found in
those evolving from $s_j$.   

\item For each $g\in \mbox{Pre}(s)$ and a child $g'$ of $g$, if $g'\not\in
\mbox{Pre}(s)$, there exist $s_j$ such that  $g'$ is mapped to $s_j$ or a node below it.
\end{itemize}

To resolve the non-binary node $s$, we need to replace the
 star tree consisting of
$s$ and its children with  a rooted binary tree $T_s$ with root $s$
and $n(s)$ leaves each labeled by a unique  $s_i$, $1\leq i\leq n(s)$.
It is well known that  $T_s$ has an equivalent partial
partition system
\begin{eqnarray*}
{\mathcal P}(T_s)&=&\{ [L(u_1), L(u_2)]: u_1 \mbox{ and } u_2 \mbox{
are siblings in } T_s\}
\end{eqnarray*}
over $\{s_1, s_2, \cdots, s_{n(s)}\}$. The partition  corresponding
to the children of the root of $T_s$ is called the {\bf first
partition}. We construct  ${\mathcal P}(T_s)$ through  computing the
first partition recursively. Therefore, we resolve $s$ by recursively solving
the so-called minimum duplication bipartition problem (Ourangraoua \textit{et
al.}, 2011). 
%
 We take this approach for
two purposes. First, it may reduce the overall duplication cost.
Second, pushing duplication down in the species tree can also reduce
the gene loss cost even if the resulting reconciliation
 is not optimal in terms of the duplication cost.

Consider a binary refinement $T_s$ of $s$. By definition, it is 
a binary tree over $s_i$ ($1\leq i\leq n(s)$). Let its first partition be
  $[A, B]$, which is the partition of the set $\{s_1, s_2, \cdots, s_{n(s)}\}$.
 For  a gene tree node $g\in \mbox{Pre}(s)$ with two
children $g_i$ ($1\leq i\leq 2$),  $g$ is associated with a
duplication occurring before the root of the refinement
  if and only if $L(g_i)\cap A\neq \phi $ and $L(g_i)\cap B\neq \phi$
for some $i$. Hence, $g$ is not associated with a duplication
occurring before the root of $T_s$ (or before $s$ in $S$) if and
only if $g$ is mapped to a node below the root or $g$ is mapped to
the root, but its children are mapped below the root. If the former
is true, $L(g)=L(g_1)\cup L(g_2) \subseteq A$ or $L(g)\subseteq B$.
If the latter is true, $L(g_1)\subseteq A$ and $L(g_2)\subseteq B$
or vise versa. Hence, $g$ is not associated with a duplication
occurring before the root if and only if
\begin{eqnarray}
  L(g_1)\subseteq A \mbox{ or }  L(g_1)\subseteq B, \label{Eq1}
\end{eqnarray}
and
\begin{eqnarray}
 L(g_2)\subseteq A \mbox{ or } L(g_2)\subseteq B. \label{Eq2}
\end{eqnarray}
The last  statement can also be generalized to non-binary gene tree
nodes. In the rest of this discussion,  for clearance, we call
$L(g_1)|L(g_2)$ a split rather than a partial partition.

Motivated by this fact, we propose to find the first partition that
maximizes the splits $L(g_1)|L(g_2)|\cdots|L(g_k)$ that satisfy the
generalization of the conditions Eqn.~(\ref{Eq1})-(\ref{Eq2}), where
the nodes $g_i$ are the children of some internal node in the gene
tree.  Formally, for a partial partition $[P,Q]$, we say that it
does not cut a multiple split $A_1|A_2|\cdots|A_k$ in the gene tree 
 if and only if for every $i$,
\begin{eqnarray}
 A_i\cap P =\phi, \;\; \mbox{ or } A_i\cap Q=\phi. \label{cut_condition}
\end{eqnarray}
 The algorithm
for finding the first partition is summarized below.
Recall that  we refine a non-binary node $s$ and its children by recursively
calling  the  first partition algorithm. \vspace{2em}

{\tt
\begin{table}[!th]
\begin{center}
\resizebox{0.95\columnwidth}{!}{
\begin{tabular}{l}
\hline\\
\hspace{3em}{\bf First Partition Algorithm} \\
\\
\;\;${\mathcal S}=\phi$; /* It is used to keep partitions */\\
\;\;For each $i$\\
\hspace{2em}FirstExtension($[\{i\}, \phi]$, $\mathcal S$);\\
\;\;Output the best partition in $\mathcal S$;\\
\\
\\
\; {\bf FirstExtension}($[P, \phi]$, $\mathcal S$) {\bf $\{$}\\
\hspace{1.5em}1. For each $i\not\in P$\\
\hspace{3em}Compute $n(i)$, the \# of the gene tree splits
 not cut by $[P, \{i\}]$; \\
\hspace{1.5em}2. Select $j$ such that $n(j)=\max_i n(i)$; \\
\hspace{1.5em}3. {\bf If} $P\cup \{j\}\neq L(S)$  {\bf do} \{\\
\hspace{3em}SplitExtension($[P, \{j\}]$, ${\mathcal S}$);
   FirstExtension($[\{j\}\cup P, \phi]$, ${\mathcal S}$);
\\
\hspace{2em}{\bf $\}$ else} \\
\hspace{3em}Add $[P, \{j\}]$ into $\mathcal S$;\\
\;$\}$ /* End of FirstExtension */
\\
\\
\;{\bf SplitExtension}($[P,Q]$, $\mathcal S$) ${\bf \{}$\\
\hspace{1.5em}1. For each $i\not\in P\cup Q$\\
\hspace{4em}Compute $n_1(i)$, the \# of the gene tree splits not cut
        by $[P, Q\cup \{i\}]$;\\
\hspace{4em}Compute $n_2(i)$, the \# of the gene tree splits not cut
         by $[P\cup \{i\},Q]$;\\
\hspace{1.5em}2. Select $j$ such that  $\max \{n_1(j), n_2(j)\}=
\max_i \{n_1(i), n_2(i)\}$;\\
\hspace{1.5em}3. {\bf If} ($P\cup \{j\} \neq L(S)$) {\bf do} $\{$ \\
\hspace{4em}SplitExtension($[\{j\}\cup P, Q]$, $\mathcal S$) if $n_1(j)\geq n_2(j)$;\\
 \hspace{4em}SplitExtension($[P, Q\cup \{j\}]$, $\mathcal S$) if $n_2(j) > n_1(j)$;\\
\hspace{2.5em}$\}$ {\bf else} $\{$\\
\hspace{4em}Add $[\{j\}\cup P, Q]$ into $\mathcal S$ if $n_1(j)\geq n_2(j)$;\\
\hspace{4em}Add $[P, Q\cup \{j\}]$ into $\mathcal S$ if $n_2(j) > n_1(j)$;\\
\hspace{2.5em}{\bf $\}$}\\
\;$\}$ /* End of SplitExtension */\\
\hline
\end{tabular}}
\end{center}
\end{table}
} \vspace{1em}

\begin{figure}[!th]
\begin{center}
\includegraphics[width=0.95\columnwidth]{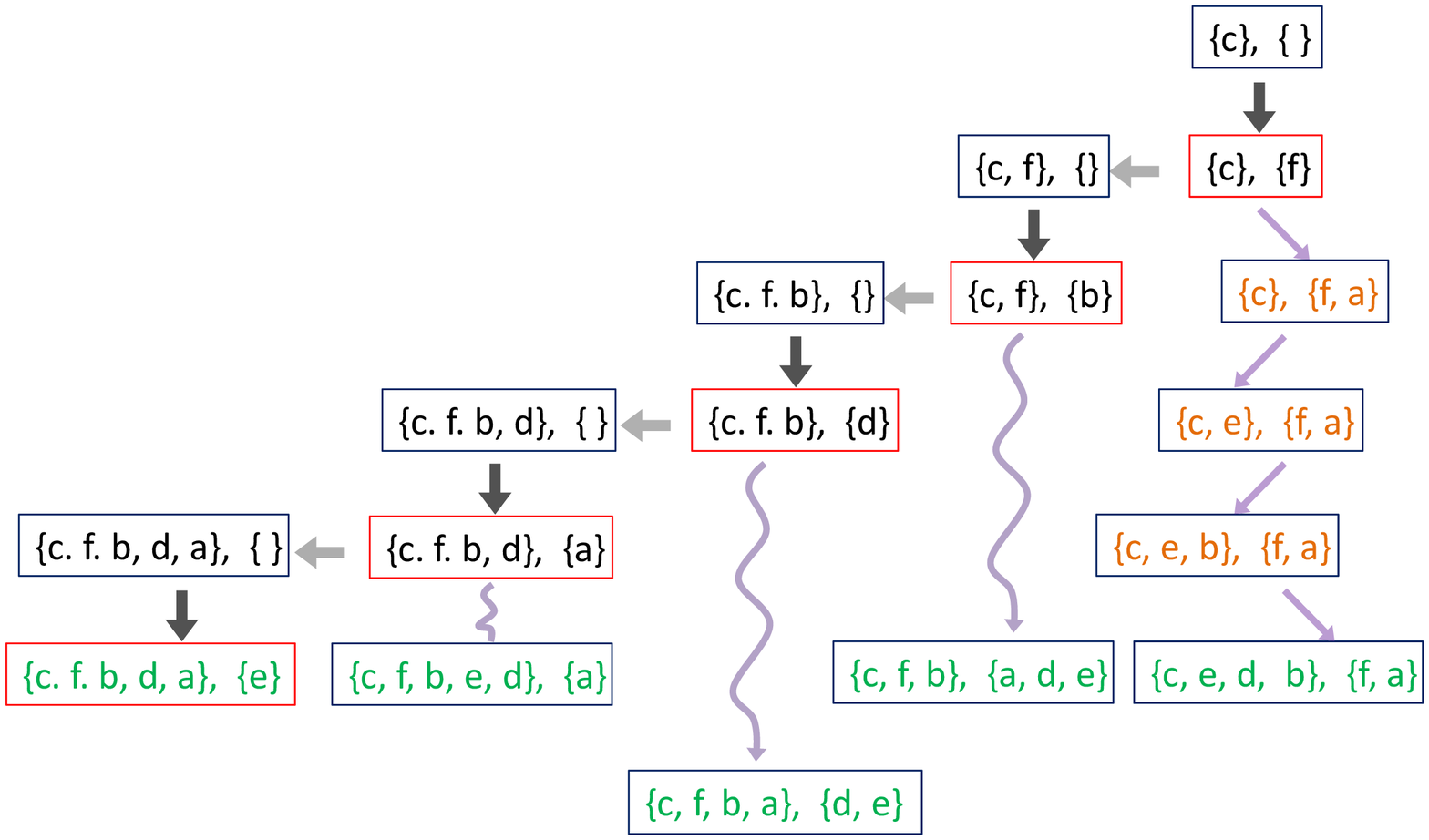}
\end{center}
\caption{Illustration of the execution of the
FirstExtension($\{c\}$, $\phi$). Here,  the considered non-binary
species tree node has children $a, b,c, d, e, f$. The subprocedure
FirstExtension() is recursively executed five times, generating
partial partitions (in red) $[\{c\}, \{f\}]$, $[\{c, f\}, \{b\}]$,
$[\{c, f, b\}, \{d\}]$, $[\{c, f, b, d\}, \{a\}]$,  and $[\{c, f, b,
d, a \}, \{e\}]$, respectively.  The SplitExtension() is called on
each of these partial partitions to produce the five partitions
shown in green in bottom. Here, the geen tree information is omitted.}
\label{Fig31_Partition}
\end{figure}

The First Partition (FP) algorithm is illustrated with an example  in
Fig.~\ref{Fig31_Partition}, where the computation flow of
the subprocedure FirstExtension($\{c\}$, $\phi$) is outlined. In
this example, we try to resolving a non-binary species tree node with
 six children $a, b, c, d, e, f$ using the splits in the gene tree. 
The gene tree splits are used in the step 1 of both FirstExtension( ) and
SplitExtension( ) and not listed explicitly  here. 
 After partial partition $[\{c\},
\{f\}]$ is obtained,  the SplitExtension( ) is called to extend
$[\{c\}, \{f\}]$ into a partition $[\{c,e, b,d\}, \{f, a\}]$ of the
child set. Since the computation of  the FirstExtension() is
heuristic, the partition  $[\{c,e, b,d\}, \{f, a\}]$ expanded from
 $[\{c\}, \{f\}]$ might not be
the optimal first partition of the child set and hence the
FirstExtension() is called on $[\{a, f\}, \phi]$ to obtain better
partitions in the case that $[\{c\}, \{f\}]$ does not lead  to the
optimal first partition. By the same reason, the
FirstExtension() is recursively called during computation. Overall,
the subprocedure FirstExtension() is recursively called five times,
outputing  the following partial partitions (in red box in
Fig.~\ref{Fig31_Partition}):
\begin{eqnarray*}
&& [\{c\}, \{f\}], [\{c, f\}, \{b\}], [\{c, f, b\}, \{d\}],\\
&& [\{c, f, b, d\}, \{a\}], [\{c, f, b, d, a \}, \{e\}];
 \end{eqnarray*}
  and the
SplitExtension() is called on these partial partitions to produce
the five partitions listed in the bottom (in green).
Then, the algorithm selects the best from these obtained partitions.

In general, assume the non-binary species tree node $s$ under
consideration has $n(s)$ children and  $k'$ gene tree nodes are
mapped to $s$. The FP algorithm calls recursively the
FirstExtension( ) $n(s)-1$ times. During each call of
FirstExtension( ), a partition candidate is generated by calling the
SplitExtension( ). When the SplitExtension() is executed,  whether a
split associated with a gene tree node is cut by a partial partition
or not is determined by verifying Eqn.~(\ref{cut_condition}) with at most $O(k')$ set operations.
Since  the
SplitExtension is recursively called at most $n(s)$ times,
the First Partition algorithm has time complexity $O(n(s)^2k')$.
Since $n(s)$ is usually small,  the algorithm runs fast.

The performance of the FP  algorithm is  evaluated  on randomly
generated data and summarized in Table~\ref{Table1}. 
Our simulation has two parameters: $c$, the number of the leaf species  below 
 the non-binary species tree node to be resolved, and $c_s$, the number of splits 
found in the gene tree. We considered eight combinations of $c$ and $c_s$.
For each combination, we generated 1000 datasets,  giving
8000 datasets in total.
 For each dataset, we ran the FP method and checkted  if 
it outputted a partition that has the maximum number
 of non-cut splits or not. 
Here, the  maximum number of  splits not cut by an optimal partition
was obtained by exhaustive search for each dataset.
 We also compared the FP algorithm
with another reported in (Ouangraoua \textit{et al.}, 2011). It is
based on an algorithm for the unweighted hypergraph min cut problem
in (Mak, 2011) and can be used for the same  purpose.
 We call it the HC algorithm.
Our tests indicate that the FP algorithm outperforms the HC
algorithm usually.

\begin{table}[!h]
\processtable{Performance of the First Partition (FP) algorithm and
an algorithm presented in (Ouangraoua \textit{et al.}, 2011).
One thousand random datasets were generated  for each combination of $c$ and $c_s$,
  which are the number of leaf species below the non-binary species tree node
to be refined  and  the number of
 splits found in the input gene tree, respectively.
 An algorithm made an error
  if it did not output an optimal partition that induces the smallest number of
first duplications.
 An entry in the last two columns 
indicates how many times the corresponding algorithm did not output an optimal
partition in 1000 tests.
\label{Table1}}
 {
\begin{tabular*}{0.9\columnwidth}{@{\extracolsep{\fill}}c
@{\extracolsep{\fill}}c @{\extracolsep{\fill}}c
@{\extracolsep{\fill}}c } \hline
  \begin{tabular}{l}
  \# of\\ elements ($c$)
 \end{tabular} &
 \begin{tabular}{l} \# of\\
splits ($c_s$)
 \end{tabular}
 & \begin{tabular}{l}
                                     \# of errors  \\
                                     for FP
                                   \end{tabular}
                               &  \begin{tabular}{l}
                                  $\#$ of errors\\
                                    for HC
                                   \end{tabular} \\
\hline \hline
   5  &  5  & 7 &  15\\ 
      & 10  & 0 &  18\\ 
   10  &  5  & 0 &  4\\ 
     &  10  & 1 &  2 \\
     &  20  & 0 &  0 \\
   15  &  7  & 0 &  3 \\
     &  15  & 0 &  1  \\
     &  30  & 0 &  1 \\
\hline
\end{tabular*}
}{}
\end{table}

Putting all the refinements at non-binary species tree nodes
together, we obtain a binary refinement $\hat{S}$ of the species
tree.

\subsection{Step Two: Resolve Non-binary Gene Tree Nodes}

 When the second step starts, a binary refinement $\hat{S}$ of the species
 tree $S$ has been obtained. In the second step,  our goal is to find
 a binary refinement $\hat{G}$ of $G$ by resolving every non-binary node in
  $G$ using $\hat{S}$ such that $\hat{G}$ has the smallest duplication cost when $\hat{G}$ and
$\hat{S}$ are reconciled. 
Moreover, the reconciliation of $\hat{G}$ and $\hat{S}$ also has the optimal
loss cost over all the reconciliations with the optimal duplication cost 
(Theorem~\ref{Dup-Loss-Proof}).
In the rest of this subsection, we present a linear time algorithm for this step.

\begin{figure}[!th]
\begin{center}
\includegraphics[width=0.8\columnwidth]{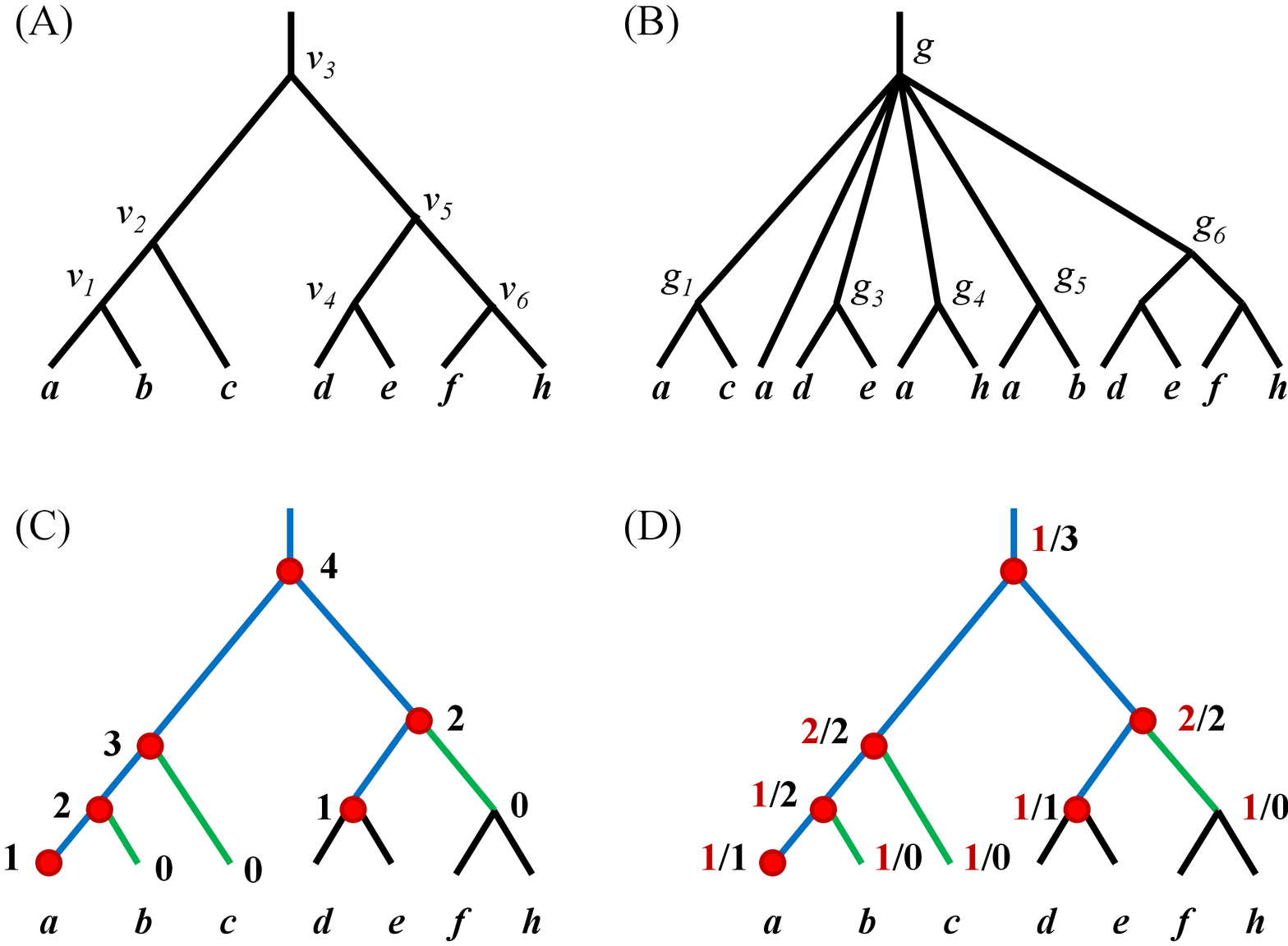}
\end{center}
\caption{An example of computing $m(\;), \alpha(\;), \beta(\;)$ for
 a gene tree and a species tree.
 ({\bf A}) A binary species tree $\hat{S}$ over 7 species $a, b, c, d, e, f, h$. 
({\bf B}) A gene tree $G$  with a non-binary root $g$.
 ({\bf C}) The subtree $I(g)$ (drawn in blue) and $I^{+}(g)$ of $\hat{S}$
 in which the number $m(x)$ written beside each node $x$.  
 The lca reconciliation map $\lambda$ from $G$ to $\hat{S}$ maps $g_1$ 
to $v_2$, $g_2$(which is a leaf) to the left child of $v_1$, $g_3$ to $v_4$, 
$g_4$ to $v_3$, $g_5$ to $v_1$, and $g_6$ to $v_5$, respectively. $I(g)$ contains
$v_i$($1\leq i\leq 5$) and the left child of $v_1$, which are highlighted in
red dot. $I^{+}(g)$ is obtained from
$I(g)$ by adding the right child of $v_1, v_2$ and $v_5$.
The edges in $I^{+}(g)$ but not in $I(v)$ are in green.
 (D) The  $\alpha(u)$ and $\beta(u)$ are
 given in the format of $\alpha(u)$/$\beta(u)$ for each $u$, from which
 three duplications and three gene losses are inferred for refining the
 non-binary node $g$.}
\label{Fig22}
\end{figure}

 We shall refine each non-binary internal node in $G$
 separately using
the lca reconciliation map $\lambda$ from $G$ to $\hat{S}$ 
 and then combine all the binary refinements to obtain $\hat{G}$. 
Consider a non-binary  internal node  $g$ in $G$. 
Let $g$ have  $k$ children $g_1, g_2, \cdots,
g_k$, where $k\geq 3$.   We first set
 $$I(g)=\{ s : \lambda(g_i)\leq s \leq \lambda(g) \mbox{ for some } i \}.$$
Note that $I(g)$ is a subset of nodes in $\hat{S}$. Furthermore,
 $I(g)$ forms
a subtree rooted at $\lambda(g)$ as shown in Fig.~\ref{Fig22}C.
 For simplicity, we also use $I(g)$ to represent the
resulting subtree.  It is easy to see that in $I(g)$ each leaf 
 is the image of some $g_i$ under $\lambda$.
 However, $I(g)$ may not be a binary subtree
because  some internal nodes may  have a child not belonging to $I(g)$ as shown in
Fig.~\ref{Fig22}C. We use
$I^{+}(g)$ to denote the binary tree obtained by including all  the
children of the non-leaf nodes of $I(g)$. 
For each species tree node $x$ in the
subtree $I^{+}(g)$, 
we define $\omega(x)$ to be the number of children that
are mapped to $x$ under the lca reconciliation $\lambda$.
 We further define $m(x)$ for each $x\in I^{+}(g)$ as

    \begin{eqnarray*}
     m(x)=\left\{ \begin{array}{ll}
        \omega(x) & \mbox{if $x$ is a leaf of $I^{+}(g)$,}\\
        \omega(x) + \max (m(x_1), m(x_2)) &
    \mbox{otherwise,}
         \end{array}
      \right.
 \end{eqnarray*}
where $x_1$ and $x_2$ are the children of $x$ if $x$ a non-leaf node
of $I^{+}(g)$,
  a subtree of $\hat{S}$. The computation of $m(\;)$ is illustrated in
Fig.~\ref{Fig22}C.

\begin{theorem}
\label{Thm22}
  At least  $m(\lambda(g))-1$ duplications are required to
produce the ancestral genes represented by $g_1, g_2, \cdots, g_k$ 
\end{theorem}
{\bf Proof.}
  Consider the partial order set (poset)
  $${\mathcal O}=(\{ L(\lambda(g_i)): 1\leq i\leq k\},
\subseteq),$$
in which an element corresponds to the image of some child of $g$ and
the binary relation is subset inclusion.
 Clearly, $m(\lambda(g))$ is the size of
the longest chain in $\mathcal O$. A subset
${\mathcal A}$ of $\mathcal O$ is an  antichain if for any $x, y\in
{\mathcal A}$, $x$ and $y$ are not comparable,
 i.e., $x \not\subseteq y$ and
 $y\not\subseteq x$.  For any $i\neq j$,  if $L(\lambda(g_i))$ and
$L(\lambda(g_j))$ are not comparable, they are disjoint since they
correspond to two  different nodes of $I^{+}(g)$, a subtree of the species tree.
 Hence,  an antichain consists of
disjoint elements in $\mathcal O$.
 Let $M$ be the smallest number of
antichains into which $\mathcal O$ may be partitioned. In (Berglund
\textit{et al.}, 2006) (see also (Chang and Eulenstein, 2005)),
 it is proved that $M-1$ is a lower bound on the
number of duplications needed to produce $g_1, g_2, \cdots, g_k$. By
a dual of Dilworth's theorem  (Mirsky, 1971), $M$ is equal to
$m(\lambda(g))$,  the size of the longest chain.    $\Box$
\vspace{0.5em}

\begin{figure}[!th]
\begin{center}
\includegraphics[width=0.7\columnwidth]{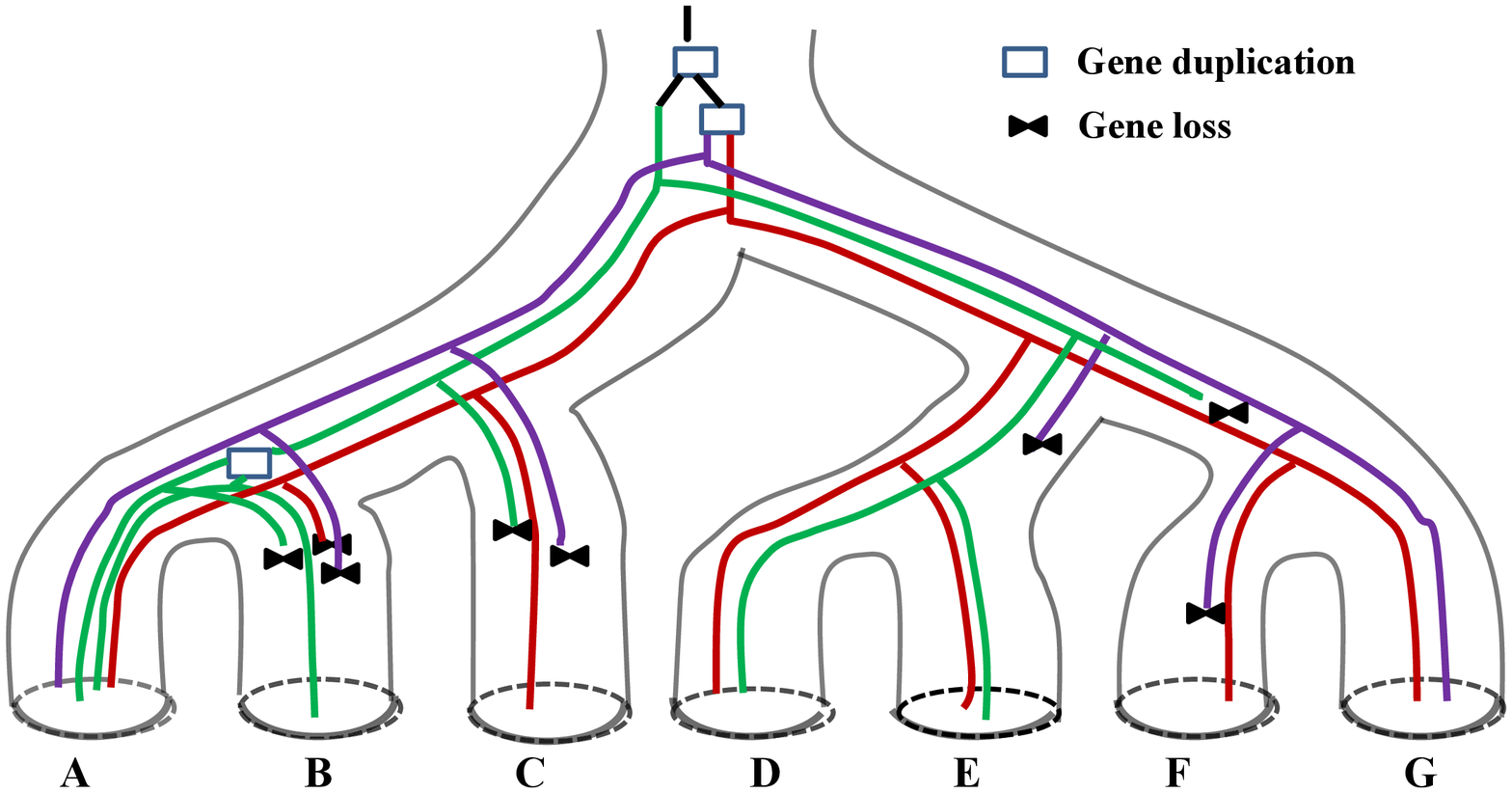}
\end{center}
\caption{
 A schematic view of the inferred evolution of  the gene family in the containing  species
tree in the example given in Fig.~\ref{Fig22}. 
({\bf A})  The binary refinement of the gene tree obtained from resolving 
the non-binary
root $g$.  
({\bf B})  A `full' reconciliation of the gene
tree and species tree, which is obtained from reconciling the obtained binary
refinement of the binary refinement of the
gene tree (in (A)) and the given species tree.}
\label{Fig1}
\end{figure}

Consider a hypothetical  evolution of a gene family in the containing
specie tree as shown in Fig.~\ref{Fig1}. In the species tree,
branches represent species. There are two numbers  
associated with each branch $e$ from $p(u)$ to $u$: 
the number $\alpha(u)$ of ancestral genes residing in the species represented by $e$ 
when it just emerged, and  the number $\beta(u)$ of ancestral genes in the
species just  before it speciated into its child species.
Clearly,
 if duplication occurred in the species, $\beta(u) > \alpha(u)$ and
 their difference
is the number of the duplication events that ever occurred, where we assume a
duplication event produced one extra gene copy; if there were gene
losses, $\alpha(u) > \beta(u)$ and their difference is the number of gene losses. 
It is easy to see that the values of $\alpha(u)$ and $\beta(u)$ are uniquely
determined by the evolution itself. Conversely, each set of such numbers
determines uniquely a family of evolutionary histories
having the same number of duplication and gene loss events. In the rest of this
section, we shall work on these numbers of a partial evolutionary
history instead of the evolutionary history itself.

 We shall infer  a reconciliation with exactly $m(\lambda(g))-1$ duplications
 associated with $g$. By Theorem~\ref{Thm22}, such a reconciliation has the
least duplication events.
The inferred duplications are postulated on the different branches
of $I^{+}(g)$ to minimize gene losses. To infer these duplications,
we define $\alpha(u)$ and $\beta(u)$ for each node $u$ of
$I^{+}(g)$ as follows. 
 Because we are working on a partial 
evolution of the gene family, $\alpha(u)$ and $\beta(p(u))$ are not
always equal, but satisfy Eqn.~(\ref{eqn:alpha_non-root}) instead.

For the root $r$ of $I^{+}(g)$,
\begin{eqnarray}
 && \hspace*{1em}\alpha(r)=1, \label{eqn:alpha_root}\\
 && \hspace*{1em}\beta(r)=\max \{\min\{m(r_1), m(r_2)\}, 1\}+\omega(r), \label{eqn:beta_root}
\end{eqnarray}
where $r_1$ and $r_2$ are the children of $r$. In general,  for a
non-root internal node $u$ with parent $p(u)$, a sibling $u'$, and
children $u_1$ and $u_2$, we have
 \begin{eqnarray}
 & & \alpha(u)=\beta(p(u))-\omega(p(u)), \label{eqn:alpha_non-root} \\
 & & \beta (u)= \left\{ \begin{array}{ll}
              m(u), \;\;\;\;\;\; \mbox{if } \alpha (u) \geq  m(u) \mbox{ or $u$ is a leaf},\\
    \gamma(u), \;\;\;\;\; \mbox{otherwise.} \\
                 \end{array}                   \right.    \label{eqn:beta_nonroot}
\end{eqnarray}
where we define
$$\gamma(u)=\max\{\alpha(u),
    \min\{m(u_1), m(u_2)\}+\omega(u), 1+\omega(u)\}.$$
 For the example in Fig.~\ref{Fig22},  the computation of $\alpha( )$ and $\beta( )$ is shown in
Fig.~\ref{Fig22}~(D).


 If $\alpha(u) < \beta(u)$, we postulate
$\beta(u)-\alpha(u)$ duplications in the branch entering  $u$; if
$\alpha(u) > \beta(u)$, we postulate $\alpha(u)- \beta(u)$ gene
losses in the corresponding branch. In total, we postulate $
\sum_{u\in I^{+}(g)}\max (\beta(u)-\alpha(u), 0)$ duplications and
$\sum_{u\in I^{+}(g)}\max (\alpha(u)-\beta(u), 0)$ gene losses.

For
the example given in Fig.~\ref{Fig22}, we infer 
two duplications above the root of the species tree and 
one duplication in the branch from $v_2$ to $v_1$ to refine the non-binary root $g$ of the 
gene tree, resulting in the binary refinement in Fig.~\ref{Fig1}A.
 The  full reconciliation of the gene tree
and the species tree given in Fig.~\ref{Fig22} can be obtain by combining 
the refinement of
non-binary root $g$ and inferences at other binary nodes and is
shown in Fig.~\ref{Fig1}B.


\begin{theorem}
\label{Dup-Loss-Proof}
 (1) The reconciliation described above requires the least
 duplications (which is $m(\lambda(g))-1$) for resolving a non-binary node $g$.

 (2) It also has the minimum loss cost over all the reconciliations with
the optimal duplication cost for resolving $g$.
\end{theorem}

The full proof of Theorem~\ref{Dup-Loss-Proof} is sophisticated and
appears in Section B of the supplementary document. However, its
idea is clear. Recall that, the non-binary node $g$ is mapped to the
root of $I^{+}(g)$. In the subtree $I^{+}(g)$, by the definition of
$m(\;)$, any path from the root $\lambda(g)$ to a leaf contains at
most $m(\lambda(g))$ images of the children of $g$; furthermore,
there is such a path $P$ containing exactly $m(\lambda(g))$ children
images. By calculating $\alpha(u)$ and $\beta(u)$ with formulas
(\ref{eqn:alpha_root})-(\ref{eqn:beta_nonroot}), we pushdown
duplications  from the root as far  as possible by postulating a
duplication in a branch of $P$ whenever it is necessary. By doing so,
 we guarantee that the resulting reconciliation has the least
gene loss cost while keeping the duplication cost unchanged. For the
example given in Fig.~\ref{Fig22}, $P$ is the leftmost  path from
the root to the leaf labeled with $a$ in the species tree. We postulate all three
duplications along $P$ and three losses off $P$.

  By preprocessing the lca map and the species
  tree $\hat{S}$, we can resolve all the non-binary gene tree nodes
  in linear time. The detail of linear-time implementation is omitted here.

\section{Implementation and Performance Analysis}

The algorithms presented above have been implemented in Python.
Given an arbitrary rooted gene (family) tree  and an arbitrary
rooted species tree, which can be binary or non-binary, our
reconciliation program outputs a
 hypothetical duplication history of the gene family.
 Although our program is  heuristic,  it usually  outputs an  evolutionary history
having the smallest user-selected reconciliation cost.
 Our program has the following features.

 \begin{enumerate}
 \item
 Following (Vernot \textit{et~al.}, 2008), our program 
 indicates whether an inferred duplication is required or
 weakly-supported.

\item For a large gene family, our program may output a set of solutions
with the same reconciliation cost.



\item Our program can take a set of arbitrary gene trees and a
species tree as its input.
When the input includes $k$ gene
trees $G_i$ ($1\leq i\leq k$) and a species tree, the program
attempts to refine all the gene trees and the species tree to 
minimize the sum of  the reconciliation costs $c(S, G_i)$, 
where $c$ is the user-selected cost function.

 Recall that a star tree is a rooted tree
in which all the leaves are the children of the root and hence
 any binary tree is a binary refinement of the star
tree over the same set of species.  
 Accordingly, our program can be used as a tool for
inferring species tree from a set of gene trees if  the star tree over the
containing species and the set of gene trees are used as input. 
The performance of our program for species tree inference is assessed in
Section~\ref{sec:test_part2}.

\item Our program can be executed from command line to allow for
automated analysis of a large number of gene trees.
\end{enumerate}

\subsection{Validation Test I: Inferring Tor Gene Duplications}

The target of rapamycin (Tor) gene is responsible for nutrient-sensing and 
 highly conserved among eukaryotes.
 In mammals, the unique mTor governs
cellular processes via two distinct complexes Tor Complex1 (TorC1)
and TorC2. However, in the budding yeast S. \emph{cerevisiae}, the
fission yeast S. \emph{pombe}, and other fungal species, there are
two Tor paralogs. Moreover,  four Tor paralogs have been found in
\emph{Leishmania major} and \emph{Trypanosoma brucei}, two species
of  phylum Kinetoplasta (Kinetoplastids).

\begin{figure}[!tbh]
\begin{center}
\includegraphics[width=0.95\columnwidth]{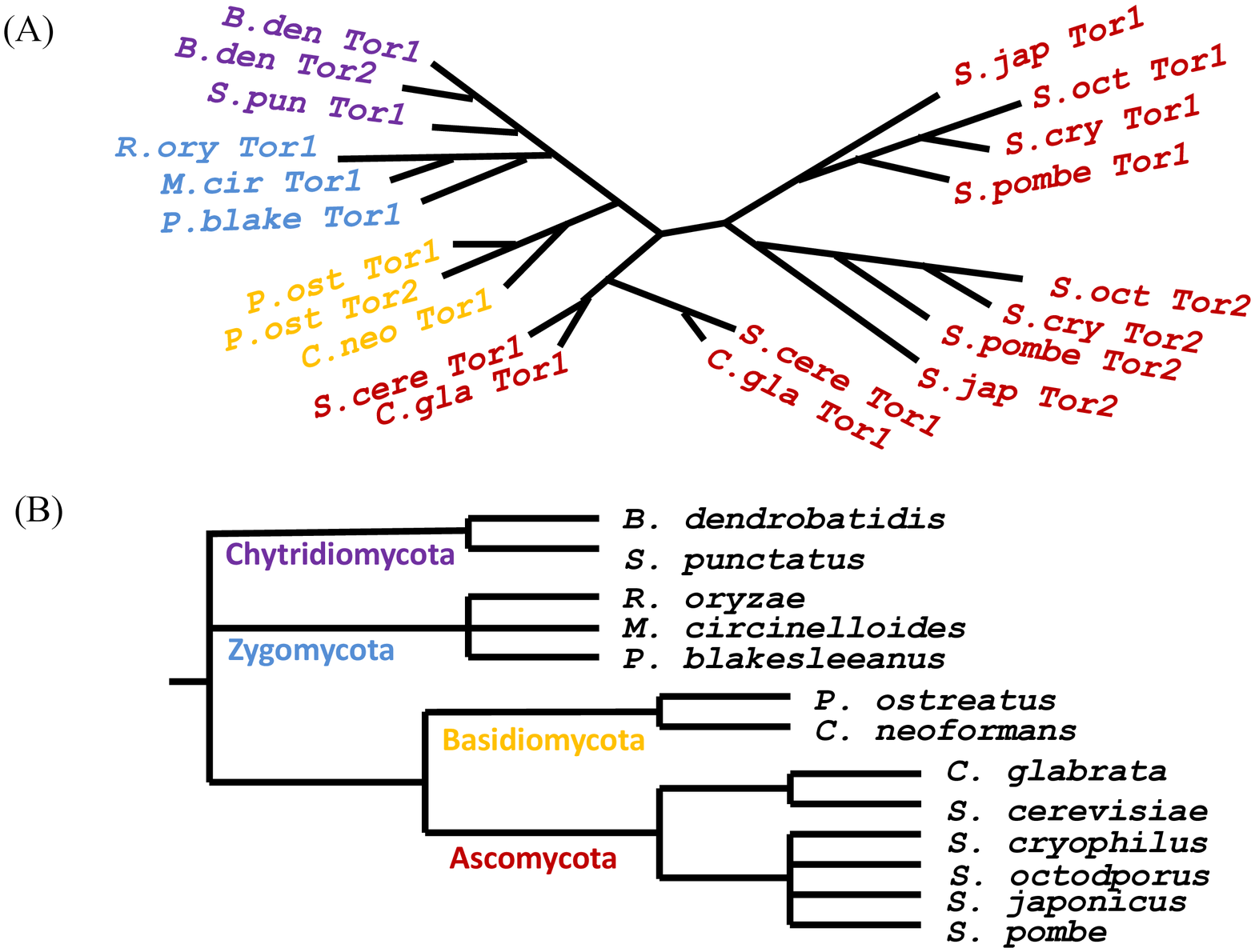}
\end{center}
\caption{(A) A Tor gene tree over thirteen fungal species,
redrawn based on the phylogenetic relationship of the Tor genes
reported in (Shertz \emph{et~al.}, 2011). (B) A non-binary species
tree of the studied fungal species downloaded from the NCBI taxonomy
database.}
\label{TorGeneTree}
\end{figure}

Shertz \textit{et~al.} (2011) investigated the evolution of the Tor family in the
fungal kingdom. They reconstructed the Tor tree over thirteen fungal
species (redrawn in Fig.~\ref{TorGeneTree}A) and from it inferred
four duplication events that are responsible for producing two Tor
paralogs in fungal kingdom. A whole genome duplication (WGD) event
is inferred, occuring  in the ancestor of \emph{S. cerevisiae} approximately one
hundred million years ago;  \emph{S. cerevisiae, S.~paradoxus}, and
other species that descend from the ancestor retained two Tor
paralogs. However, three independent lineage-specific duplications
are responsible for the two paralogs in \emph{S. pombe}, \emph{B.
dendrobatids} and  \emph{P. ostreatus}, respectively. When we
applied out program to the Tor tree and the non-binary species tree
downloaded from the NCBI taxonomy database (drawn in
Fig.~\ref{TorGeneTree}B), the same set of duplications were
inferred.

\subsection{Validation Test II: Gene Guplications in \emph{Drosophila}}
\label{sec:test_part2}

\begin{figure}[!tbh]
\begin{center}
\includegraphics[width=\columnwidth]{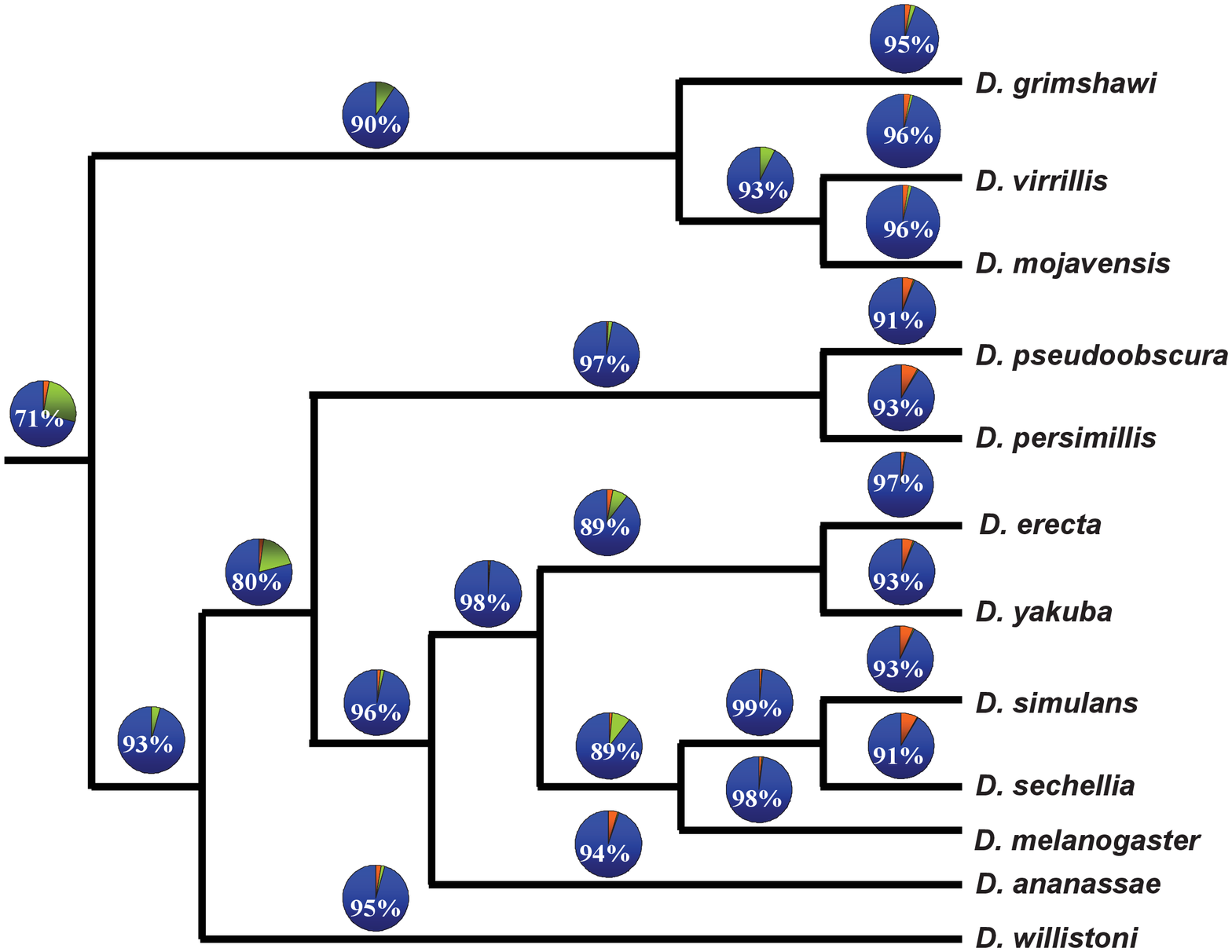}
\end{center}
\caption{Comparison of our method and CAFE (Hahn
\textit{et~al.}, 2005) on the {\it Drosophila} gene families. The
branch lengths are arbitrary in the species tree. In a pie chart,
the three sectors represent the proportions
of multiple gene families for which both methods infer same
duplications (blue, also given in percentage), only CAFE
inferred duplications (orange) and only our method inferred duplications
(shallow green), respectively.}
\label{FlyAccuracy}
\end{figure}

We further apply our reconciliation program to study gene duplication
in  the \emph{Drosophila} species.  We used  the gene tree data 
prepared by Hahn (2007). It contains 13376 gene trees over twelve \emph{Drosophila} species.
 The 3707 of the gene families contain multiple gene instances in at least one 
species,  whereas the rest are single-gene families. We compared our
program with CAFE, a statistical program for duplication inference
reported in (Hahn \textit{et~it.}, 2005) on the multiple gene families.
For each multiple gene family, we first
contracted edges having low support value in each  gene (family) tree
using cut-off value $X$(80, 90, or 100) and ran our
program on the resulting gene trees, which may or may not be binary.
Our program had similar performance for the three cut-off values.
Fig.~\ref{FlyAccuracy} shows the performance of our program
when the cut-off value is set to 80. 

 We also ran CAFE for the multiple-gene families. 
Since the duplication inference of
CAFE is independent of the family gene tree, the cut-off value used
for processing gene trees has no impact on CAFE's performance.

Except for the root branch
and three others, both programs identified the same duplication
events for over 90\% of multiple gene families. Clearly,  our method
inferred more duplication events along deep branches, whereas CAFE
inferred more along branches ending with a leaf, called {\it informative
branches},  consistent with the
observation made by Hahn (2007). In fact, CAFE often overestimates
duplications in the informative branches in our
simulation test on the same species tree reported in
Section\ref{sec:test_part3}. Hence, combining the both methods should give
accurate  estimation of the gene duplications occurring on both 
deep and informative branches
 in the species tree.

\begin{table}[!hb]
\processtable{Accuracy of inferring the unrooted \emph{Drosophila}
species tree form unrooted gene trees. Accuracy0: The accuracy of
inferring the species tree from original gene trees obtained in
(Hahn, 2007); accuracyX: The accuracy of the inference with the
non-binary gene trees obtained from the original gene trees via
branch contraction with the cut-off value X=60, 90.
 \label{table22}} {\begin{tabular}{cccc}
\hline No. of gene trees &
Accuracy0(\%) & Accuracy60(\%) & Accuracy90(\%)\\
\hline
   5    &21   &35   &34\\
  10    &45   &72   &54\\
  20    &61   &87   &68\\
  30    &76   &92   &84\\
\hline
\end{tabular}}{}
\end{table}

When a set of unrooted gene trees (and a  star tree) are used as
input, our program infers an unrooted binary species tree. We used  the
\emph{Drosophila} gene trees to test our program in inferring
unrooted species tree. We used the original gene trees and the
classes of non-binary gene trees obtained from branch contraction
with cut-off value 60 and 90. From the results given in
Table~\ref{table22}, we observe that contracting weakly supported
edge (with support value below 60\%) improves greatly the accuracy
of inferring unrooted species tree.  It is also true that contracting high-supported
branches reduces the accuracy of inferring species tree.

\subsection{Validation Test III: Simulation}
\label{sec:test_part3}

We assess both  the CAFE and our method for gene duplication
inference through random simulation on the same \emph{Drosophila} species tree
as used in Section~\ref{sec:test_part2}. 
The twelve species covered in the species tree have evolved
 from their least common ancestor in the
past roughly 63 million years (Hahn, 2007). 
We generated 1000 random gene families  in the birth-death model by setting
 both duplication and loss rates to 0.002 per million years, which are estimated from the 
gene evolution in the species tree (Hahn \textit{et al.}, 2005). 
Each random gene family includes a small number of  instances in a species.
For each gene family, we
recorded gene duplication  and loss events occurring along every branch of the
species tree; we then derived its gene tree from the recorded duplication events.

From the true tree of a random gene family, we also derived two
approximate gene trees by contracting
 branches that are shorten than   2 and 3 million years, respectively.
The resulting trees may or may not be binary for each gene family.
We ran our program to infer duplication events by reconciling each of the three obtained
 trees and the species tree for each gene family.  
We then  computed  the accuracy of our program  for 
duplication inference  in each of the three cases. 
Recall that the CAFE program infers gene duplication events without using gene
tree information.
For each gene family, we simply ran the CAFE program  using the same duplication and loss 
rates 0.002 per million years and computed its accuracy.

The performance of the two  programs is 
summarized in a table in the Section C of the supplementary
document.
%
 As a
reconciliation method, our program uses the structural information
of a gene tree to infer gene duplication and thus tends to overestimate duplication
 events along deep branches.
In our test, it inferred correctly  the duplication history  from the true gene tree for
all except for one  gene families.
 When the trees obtained
from edge contraction were used, our program overestimated
duplications frequently. But it still has high accuracy to detect 
duplications on both deep and informative branches.
 In contrast, the CAFE program often overestimated duplications along the informative branches. 
We noticed that
it also overestimated duplications on the root branch (the first branch
in the table). The reason for this fact is   unclear.

Additionally, we  used the same simulated data to evaluate the 
accuracy of the binary refinement of the input non-binary species tree.
 Here, we
assume the species tree is correctly rooted. We contracted the
branches shorter than 10 million years in the species tree,
obtaining the following non-binary tree (in Newick format):

\noindent{\it
((dgri,dmoj,dvir),dwil,(dpse,dper),(dmel,dsec,dsim,dere,dyak,dana)).}\\
The accuracy analysis  is reported in Table~\ref{table2}. When a set of
true gene trees  was  used, the program could output the true species
tree as the binary refinement of the above non-binary species tree.
When a set of contracted gene trees was used, the program also
performed well. For example, with more than 15 gene trees derived
from contracting about 3 edges, our program could recover the true
species tree from the non-binary species tree given above with
accuracy over 97\%.

\begin{table}[!th]
\processtable{Accuracy of the binary refinement of the input
non-binary species tree. The accuracy is given in percentage of the cases
for which the program outputted the \emph{Drosophila} species tree as the
binary refinement of the non-binary input tree (over 100 tests for each entry in the table).
 $N$ is  the
number of input gene trees;  A is  the accuracy of the output binary
refinement. \label{table2}} {
\begin{tabular}{rc|r|cc} \hline
 $N$
 &
 \begin{tabular}{l}
 Contraction\\
  rate
  \end{tabular}
 &
   A(\%)
    & \begin{tabular}{l}
Mean no. \\
of removed\\
edges
\end{tabular} & \begin{tabular}{l} Max. node \\
                degree \end{tabular}\\
\hline
   2    & 0.1   &65   &1.03   &2.79\\
   5    &    &95   &0.97   &2.73\\
  10    &    & 100   &0.99   &2.75\\
  15    &    & 100   &1.03   &2.75\\
  20    &    & 100   &0.99   &2.72\\
  30    &    & 100   &0.99   &2.73\\
   2    & 0.3   &26   &2.98   &3.82\\
   5    &    &72   &2.91   &3.73\\
  10    &    & 90  &2.95   &3.78\\
  15    &    &97   &2.90   &3.75\\
  20    &    &99   &2.95   &3.77\\
  30    &    &100   &2.99   &3.80\\
   2    & 0.5   &7   &4.84   &5.03\\
   5    &    &27   &4.83   &4.96\\
  10    &    &65   &5.00   &5.14\\
  15    &    &66   &4.94   &5.09\\
  20    &    &76   &4.91   &5.01\\
  30    &    &90   &5.02   &5.08\\
\hline
\end{tabular}
}{}
\end{table}

\section{Discussion}

We have been investigated the general reconciliation problem,
in which both input gene and species trees can be non-binary. Only
special cases of this problem had been studied in literature.
 When the
input species tree is binary and the input gene tree is non-binary,
the reconciliation problem is polynomial-time solvable through a dynamic programming
approach (Chang and Eulenstein, 2006; Durand \textit{et al.}, 2005).
However, if the input species tree is non-binary,
the problem becomes much more hard. Vernot \textit{et al.} (2008) 
developed a heuristic method for this case. 

In this paper, we approach the general reconciliation problem via finding 
the binary refinements of gene tree and species tree that minimize a reconciliation cost.
Such an approach is promising as it unifies gene duplication inference  through
tree reconciliation  with inferrng species tree from gene trees.

First, we have proved that  the  general reconciliation problem is NP-hard even for the 
duplicaiton cost. This  answers an open problem on tree reconciliation (Eulenstein \textit{et al.}, 2010; Vernot \textit{et al.}, 2008). 
It suggests that the general reconciliation problem is unlikely polynomial time
solvable.

We then present a fast heuristic algorithm to solve the general
reconciliation problem. Given a gene tree $G$ and a species tree
$S$, we reconcile $G$ and $S$ in two steps. In the first step,
 a binary refinement
$\hat{S}$ of  $S$ is  computed using the structural information of $G$ if
$S$ is non-binary.
We have presented a novel algorithm for the purpose.  The algorithm for 
the minimum duplication speciation problem given in Ourangaoua \textit{et al.} (2011)
can be used in this step. 
However, our validation test shows that our proposed algorithm outperforms
 theirs. This step will not be executed if $S$ is a binary tree. 

 In the
second step, a binary refinement $\hat{G}$ of $G$ is computed using 
$\hat{S}$ if $G$ is not binary. We have developed  a linear-time
algorithm for this step. Our algorithm
benefits from an elegant theorem in order theory (Mirsky,
1971). We focus on the longest chain instead of  disjoint partitions
of the images of the children of a non-binary node in $G$
(Berglund et al., 2006; Chang and Eulenstein, 2006). Our method
outputs a reconciliation with the optimal duplication cost.
Moreover,  it has the smallest gene loss cost over all
 reconciliations with the optimal duplication cost.
When two  binary trees are reconciled, the lca
 reconciliation has not only the best duplication cost
(Gorecki and Tiuryn,  2006), but also the optimal gene loss cost
(Chauve and El-Mabrouk, 2009). However, such a reconciliation simply
does not exist for non-binary gene trees. Our proposed algorithm for
resolving non-binary gene tree nodes is identical to the standard
duplication inference procedure when applied to binary gene tree
nodes. Thus, our algorithm can be considered
 as a natural generalization of the standard reconciliation to
 non-binary gene trees. In our implemented program, the user can
 also choose the dynamic
programming algorithm proposed by Durand \textit{et al.} (2005) to refine
the non-binary gene tree in the second step.

Our
algorithm has been implemented into a computer program which is online
available to evolutionary biology community.
A tree reconciliation method often overestimates
duplication events along a deep branch in the input  species tree (Hahn, 2007).
 First, such a method takes into account both gene copies in extant
species and gene tree structure. When gene tree and the containing
species tree are inconsistent at an internal tree node, duplication
has to be assumed. Therefore, a deep coalescence  could lead to
overestimation of gene duplication events along the branch where the deep
coalescence event occurred.  However, our preliminary study 
suggests that the effect of deep coalescence on gene duplication inference is not as  severe
as previously thought.   
 Secondly, deep branches in both gene and species trees
are often reconstructed with low support value because of artifacts
caused by low taxon sampling or long branch attraction (Koonin,
2010). Any error occurring in deep branch estimation might lead to
overestimation of duplications along an incorrectly-inferred deep
branch. Our method attempts to reduce the error of the second type by
reconciling non-binary gene and species trees.

Probabilistic approaches assume that gene duplication and loss
events are neutral processes and provide a natural setting for
incorporating sequence evolution directly into the reconciliation
process (Akerborg \textit{et al.}, 2009;  Arvestad \textit{et al.},
2004; Arvestad \textit{et al.}, 2009;  Gorecki and Eulenstein,
2011), but they are computation and data intensive. Our approach is based
on parsimony principle and thus better suited to data sets where gene evolution events are
 rare. Hence, our method is  complement to the probability-model-based approach. 
For instance, the CAFE program often
overestimated duplications in informative branches, while our
program is quite accurate on them.

Finally, our method for refining non-binary species tree can
actually be used for reconstructing species trees from a set of gene
trees. Different heuristic methods for species tree inference have
been proposed recently (Than and Nakhleh, 2009; Liu and Pearl,
2007). Our experimental test indicates that our proposed method is
quite promising for this purpose.
 It is interesting to
explore our approach for species tree inference further in future.


\section*{Acknowledgment}
LX Zhang would like to thank
Daniel Huson for suggestion of working on reconciliation with non-binary trees. He would also like to thank
C. Chauve and David A. Liberles  for comments on the preliminary version of this paper.

\paragraph{Funding\textcolon} The Singapore MOE grant
R-146-000-134-112.



\end{document}